
\magnification=1200
\hsize 15true cm \hoffset=0.5true cm
\vsize 23true cm
\baselineskip=15pt

\font\medio=cmr10 scaled \magstep2
\outer\def\beginsection#1\par{\medbreak\bigskip
      \message{#1}\leftline{\bf#1}\nobreak\medskip\vskip-\parskip
      \noindent}

\def \pa {\partial}
\def \ra {\rightarrow}

\def \ti {\tilde}
\def \dag {\dagger}

\def \Da {\Delta}

\def \a {\alpha}

\def \ga {\gamma}
\def \sg {\sigma}
\def \Sg {\Sigma}
\def \da {\delta}

\def \r {\rho}
\def \om {\omega}

\def \noi {\noindent}

\def\sqr#1#2{{\vcenter{\hrule height.#2pt\hbox{\vrule width.#2pt
height#1pt \kern#1pt\vrule width.#2pt}\hrule height.#2pt}}}

\def\lsim{\mathrel{\rlap{\lower4pt\hbox{\hskip1pt$\sim$}}
    \raise1pt\hbox{$<$}}}         
\def\gsim{\mathrel{\rlap{\lower4pt\hbox{\hskip1pt$\sim$}}
    \raise1pt\hbox{$>$}}}         

\line{\hfil DFTT-63/92}
\line{\hfil October 1992}
\vskip 2 cm
\centerline{\medio ENTROPY PRODUCTION}
\vskip 0.5true cm
\centerline{\medio IN THE COSMOLOGICAL AMPLIFICATION}
\vskip 0.5 true cm
\centerline{\medio OF THE VACUUM FLUCTUATIONS}
\vskip 1cm
\centerline{M.Gasperini and M.Giovannini}
\bigskip
\centerline{\it Dipartimento di Fisica Teorica dell'Universit\`a,}
\centerline{\it Via P.Giuria 1, 10125 Torino, Italy,}
\centerline{\it and}
\centerline{\it Istituto Nazionale di Fisica Nucleare, Sezione di Torino}
\vskip 3 cm
\centerline{\medio Abstract}
\noindent
We estimate the entropy associated to a background of squeezed cosmic
gravitons, and we argue that the process of cosmological pair production
from the vacuum may explain the large amount of entropy of our present
universe.
\vskip 1 true cm
\noi
--------------------------------
\vskip 1 true cm
To appear in {\bf Phys.Lett.B}

\vfill\eject

{\bf Entropy production in the cosmological amplification of the vacuum

fluctuations}

A successful inflationary model is expected to solve the various
problems of the standard cosmological scenario and then, in particular,
must provide some explanation to the unnaturally large value of the
entropy to-day stored in the cosmic microwave background (CMB). The
solution to the entropy problem relies, conventionally, on the so-called
"reheating" process [1], occurring at the end of the inflationary era,
and realized in various model-dependent ways such as bubble collisions
and/or inflaton decay. In view of the difficulties encountered by some
model in providing an efficient enough reheating phase [1,2], however,
it seems interesting to point out an additional source of entropy in the
inflationary scenario: the parametric amplification of the vacuum
fluctuations, or (equivalently stated, in a second quantization
formalism) the pair production from the vacuum, due to changes in the
background metric field.

As discussed in previous papers [3-5], the growth of the average number
of field quanta, due to pair production, is naturally associated
to the growth of the entropy of that field. On the other hand the pairs
spontaneously emitted from the vacuum, in an evolving cosmological
background, necessarily appear in a "squeezed" quantum state [6,7],
namely in a state in which the quantum fluctuations of a given field
operator $Q$ are suppressed with respect to the vacuum, while the
fluctuations of the canonically conjugate operator $P$ (called
"superfluctuant") are correspondingly enhanced (see, for instance,
[8,9]).

In order to compute the entropy associated to cosmological pair
production, we shall thus propose in this paper a "coarse graining"
approach to non-equilibrium entropy, valid for squeezed states, in which
the loss of information associated to the reduced density matrix is
represented by the increased dispersion of the superfluctuant operator
$P$. Such a quantum-mechanical approach is different from a recent
field-theoretic computation of the entropy of the cosmological
perturbations [10,11]. The two methods provide results which coincide in
the large squeezing limit; our expression for the entropy, however, is
always positive definite (unlike the classical approach of [10]), and
seems then to be valid even in the small squeezing regime.

The definition of entropy we shall provide holds in general for any
mechanism of particle production from the vacuum (more generally, from
any coherent initial state), not necessarily of gravitational origin,
which can be represented as a mixing of positive and negative frequency
modes through an appropriate Bogoliubov transformation. We shall apply
this definition, in particular, to estimate the entropy change
associated to the formation of a background of cosmic gravitons. By
using a very general model of cosmological evolution, we shall find an
entropy which may be as large as the CMB electromagnetic entropy,
provided the transition from inflation to the radiation-dominated phase
occurs at a curvature scale of the Planck order.

Let us recall, first of all, that the quantum description [6,7] of the
amplification of scalar or tensor fluctuations is based on the
separation of the field into background solution and first order
perturbations, and on the expansion of the solution to the perturbed
wave equation into $|in>$ and $|out>$ modes. The complex coefficients of
this expansion are interpreted, in a second-quantization formalism, as
annihilation and creation operators for a particle ($b, b^\dag$) and the
corresponding anti-particle ($\ti b , \ti b ^\dag $). The relation
between $|in>$ and $|out>$ mode solutions can thus
be expressed, for each
mode $k$, as a Bogoliubov transformation between the $|in>$ operators
($b, b^\dag , \ti b , \ti b^\dag $) and the $|out>$
ones ($a, a^\dag, \ti
a , \ti a^\dag $):
$$
\eqalign
{a_k&=c_+(k)b_k + c_-^*(k) \ti b^\dag _{-k} \cr
\ti a^\dag _{-k} &=c_-(k)b_k + c_+^*(k) \ti b^\dag _{-k} \cr} \eqno(1)
$$
where $|c_+|^2-|c_-|^2=1$. By parametrizing the Bogoliubov coefficients
$c_{\pm}$ in terms of the two real numbers $r \geq 0$ and $\theta$,
$$
c_+(k)= \cosh r(k) ~~~~~,~~~~~c_-^*(k)=e^{2i\theta_k} \sinh r(k)
\eqno(2)
$$
the relations (1) can be re-written as unitary
transformations generated by the (momentum-conserving) two-mode
squeezing operator $\Sg_k$,
$$
\Sg_k= \exp(z_k^* b_k\ti b_{-k} - z_k b_k^\dag \ti b_{-k}^\dag )
{}~~~,~~~ z_k=r(k) e^{2i\theta_k} \eqno(3)
$$
($r$ is the so-called squeezing parameter) as
$$
a_k = \Sg_k b_k \Sg_k^\dag \eqno(4)
$$
(and related expressions for $\ti a^\dag, a^\dag$ and $ \ti a$).
Starting with the $|in>$ vacuum state $|0>$, such that $b_k |0>=0=
\ti b_{-k} |0>$, the pair production process leads then, for each mode,
to the "squeezed vacuum" state [6,7] $|z_k>= \Sg_k |0>$, which satisfies
$a_k |z_k>=0=\ti a_{-k} |z_k>$, and which has an averaged particle
number
$$
\overline n_k= <z_k|b_k^\dag b_k|z_k> =
<z_k|\ti b_{-k}^\dag \ti b_{-k}|z_k> =
<0|a_k^\dag a_k|0> = |c_-(k)|^2= \sinh^2 r(k)
\eqno(5)
$$

For the purpose of this paper it is convenient to note that the two-mode
transformation (4) can be factorized as the product of two one-mode
transformations, as usually done in squeezed-state theory [9], by
defining a new pair of annihilation operators ($a_1, \ti a_1$) through
the complex rotation
$$
a={1\over \sqrt 2}(a_1-i\ti a_1) ~~~~,~~~~
\ti a={1\over \sqrt 2}(a_1+i\ti a_1) \eqno(6)
$$
and a similar expression for $b, \ti b $,
$$
b={1\over \sqrt 2}(b_1-i\ti b_1) ~~~~,~~~~
\ti b={1\over \sqrt 2}(b_1+i\ti b_1) \eqno(7)
$$
(obviously, by definition, $[a_1,a_1^\dag]=[\ti a_1,\ti a_1^\dag]
=[b_1,b_1^\dag]=[\ti b_1,\ti b_1^\dag]=1$, $[a_1,\ti a_1]=
[b_1,\ti b_1]=0$). The transformation (4) can thus be rewritten, for
each mode, as
$$
a=b_1 \cosh r + b_1^\dag e^{2i\theta} \sinh r
-i (\ti b_1 \cosh r +\ti b_1^\dag e^{2i\theta} \sinh r)=
 \Sg_1 b_1 \Sg_1^\dag  -i\ti \Sg_1 \ti b_1 \ti \Sg_1^\dag
\eqno(8)
$$
where $\Sg_1$ and $\ti \Sg _1$ are the one-mode squeezing operators for
$b_1$ and $\ti b_1$,
$$
\Sg_1= \exp({z^*\over 2} b_1^2 - {z\over 2} b_1^{\dag 2})~~~,~~~
\ti \Sg_1= \exp({z^*\over 2} \ti b_1^2 - {z\over 2} \ti b_1^{\dag 2})
\eqno(9)
$$

Given any two-mode squeezed state $|z>$, with squeezing parameter $r>0$,
it is always possible to introduce two operators $x$ and $\ti x$ whose
variances $(\Da x)_z$, $(\Da \ti x)_z$ are amplified with respect to the
vacuum value $(\Da x)_0$, $(\Da \ti x)_0$, namely [8,9]
$(\Da x)_z / (\Da  x)_0 = (\Da \ti x)_z /(\Da \ti x)_0 = \exp(r)$, where
$(\Da x^2)_z$=$ <z|x^2|z>$-$(<z|x|z>)^2$ (and the same for $\ti x$). In
terms of these operators, $b_1$ and $\ti b_1$ have the differential
representation
$$
b_1={e^{i(\theta + \pi/2)}\over \sqrt 2}(x+\pa_x) ~~~~,~~~~
\ti b_1={e^{i(\theta + \pi/2)}\over \sqrt 2}(\ti x+\pa_{\ti x})
\eqno(10)
$$
where the phase has been chosen according to eq.(2), in such a way to
identify the $x$ and $\ti x$ operators with the superfluctuant ones.

Therefore, in the $(x,\ti x)$-space representation
the wave function $\psi_{z_k}$
for the two-mode squeezed vacuum, determined by the condition $a \psi_z
=0$, can  be factorized according to eq.(8) as
$$
\psi_{z_k}(x,\ti x)=<x \ti x |z_k>= \psi^1_{z_k}(x)\psi^1_{z_k}(\ti x)
\eqno(11)
$$
where $\psi^1_{z}(x)$ is fixed by the differential equation
$$
\cosh r (x+\pa_x)\psi^1=\sinh r (x-\pa_x)\psi^1 \eqno(12)
$$
With the normalization condition $<z_k|z_k>=1$ one  obtains
$$
\psi_{z_k}(x,\ti x)=({\sg_k \over \pi})^{1/2} exp [-{\sg_k \over 2}
(x^2 +\ti x^2)] \eqno(13)
$$
where the real number $\sg_k=exp[-2r(k)]$ measures the departure of
$\psi_{z_k}$ from the vacuum wave function ($\sg=1$).

The dynamical evolution leading to the formation of squeezed states
is thus accompanied, in general, by a loss of information corresponding
to a larger dispersion of $x$ and $\ti x $ around their mean values,
$\Da x= \Da \ti x = 1/\sqrt{2 \sg}$. Such an increase in uncertainty is
measured, in the $(x,\ti x)$ representation, by the "flattening" of the
gaussian (13) with respect to the vacuum wave function, with a
probability distribution $P_k (x,\ti x)=
|<x \ti x |z_k>|^2$ which defines the reduced density operator
$$
\rho_k= \int dx d\ti x |x \ti x><x \ti x |z_k><z_k|x\ti x><x \ti x |
{}~~~,~~~ Tr \r_k =1 \eqno(14)
$$
The entropy production associated to the transition $|0> \ra |z_k>$,
according to this reduction scheme, can now be computed by applying the
usual information-theoretic definition of entropy ($S=-Tr \r \ln \r$),
and by subtracting the constant contribution of the vacuum state,
$$
S_0=-{1\over \pi}\int dx d\ti x e^{-(x^2+\ti x^2)} \ln
({e^{-(x^2 +\ti x^2)} \over \pi}) = 1+\ln \pi \eqno(15)
$$
The result can be simply expressed in terms of the averaged particle
number $\overline n_k$ as
$$
S(k)= - Tr \r_k \ln \r_k - S_0=
-\int dx d\ti x P_k(x, \ti x)\ln P_k(x, \ti x) - S_0=
$$
$$
=2 r(k)=2  \sinh^{-1}|c_-(k)| = 2 \ln (\sqrt{\overline n_k}
+ \sqrt{1+\overline n_k}) \eqno(16)
$$

This expression for the entropy density per mode holds even if the
initial vacuum is replaced in general by a coherent state $|\a_k>$, such
that $b_k |\a_k>=\a_k |\a_k>$; moreover, it is always positive definite,
and can be applied for any value of the squeezing parameter $r$. In the
large $\overline n_k$ (or, equivalently, large squeezing) limit, eq.(16)
gives
$$
S(k)\simeq \ln |c_-(k)|^2 = \ln \overline n_k \eqno(17)
$$
which coincides exactly ( for $\overline n_k >>1$) with the entropy
associated to an oscillator in a state of thermal equilibrium, and with
the result of [10,11] (such a relation between the entropy and the
mixing Bogoliubov coefficient was first suggested in [3]). The same
result can be obtained if the reduction procedure, from the pure squeezed
state down to a statistical mixture, is performed (mode by mode) in the
Fock space spanned by the eigenstates $|n_k \ti n_{-k}>$ of the number
operators $N_k=b_k^\dag b_k$, $ \ti N_k= \ti b_{-k}^\dag \ti b_{-k}$.

Indeed, as stressed in [6], the number $N$ is a superfluctuant operator
in the state $|z_k>$, while the variance of the conjugate phase operator
is squeezed. For large values of $r$, the loss of information with
respect to $N$ is measured, as discussed also in [5,11], by the density
operator
$$
\r_k =\sum_n P_k(n) |n_k \ti n_{-k}><n_k \ti n_{-k}| \eqno(18)
$$
where $P_k(n)=|<n \ti n|z_k>|^2$ . It is well known, on the
other hand, that any process of pair production from the vacuum
described by the Bogoliubov transformation (1) is characterized by a
probability distribution [12]
$$
P_k(n)=
{|c_-(k)|^{2n} \over (1+|c_-(k)|^2)^{n+1}} \eqno(19)
$$
One thus obtain, for $|c_-|>>1$, that $S(k)= -Tr \r_k \ln \r_k \simeq \ln
|c_-(k)|^2$, in agreement with eq.(17).

The expression (16) provides the entropy density per mode for all cases
of particle production (not necessarily of gravitational origin)
parametrized by the Bogoliubov transformation (1). The total entropy $S$
in a proper spatial volume $V$ is thus obtained by summing the squeezing
contributions over all modes of proper frequency $\om$ (following the
coarse graining approach of [4,5], we have neglected any correlation
among modes)
$$
S=
{V\over \pi^2}\int_{\om_0}^{\om_1} r(\om) \om^2 d\om \eqno(20)
$$
where $\om_0$ and $\om_1$ delimit the frequency interval in which the
squeezing mechanism is effective.

We shall apply this result in order to estimate the entropy growth
associated to the cosmological productions of gravitons in a homogeneous
and isotropic background, whose scale factor $a(t)$ describes the
evolution from a primordial inflationary phase to the standard radiation
and matter-dominated expansion. In such case proper and comoving
frequency ($k$) are related by $\om =k/a(t)$, and the graviton spectrum
is determined by the Bogoliubov coefficient as follows [13-15] (we
follow the notations of [13])
$$
\eqalign{
|c_-(\om)|& \simeq ({\om \over \om_1})^{-|\da|}
{}~~~~~~~~~~~~~~,~~~~\om_2<\om
<\om_1 \cr
|c_-(\om)|& \simeq ({\om \over \om_1})^{-|\da|} ({\om \over \om_2})^{-1}
{}~~~~~~,~~~~\om_0<\om
<\om_2 \cr} \eqno(21)
$$
Here $\da$ is an order of unity parameter depending on the kinematical
behaviour of the inflationary phase ($|\da |=2$, for instance, in the
case of de Sitter inflation); $\om_1$ is the maximum cutoff frequency
depending on the curvature scale $H_1$ at the beginning ($t=t_1$) of the
radiation-dominated evolution; $\om_2$ is the frequency corresponding to
the radiation-matter transition ($t=t_2$); finally, $\om_0 \simeq H_0
\sim 10^{-18} sec^{-1}$ is the minimum frequency fixed by the to-day
value of the Hubble radius ($\om_1>>\om _2 >>\om _0$).

In this example $|c_-|>>1$, so that $r(\om) \simeq \ln |c_-(\om)|$. By
integrating the squeezing parameter according to eq.(20) we find that
the total entropy per comoving volume is constant, just like the black-
body entropy. By keeping only the dominant terms, we obtain for the
graviton entropy $S_g$ in a comoving volume $\ell ^3$
$$
S_g={|\da| \over (3 \pi )^2} (a \ell \om_1 )^3 \eqno(22)
$$
Neglecting numerical factors of order unity, the total entropy $(a \ell
\om_1)^3$ can be easily estimated at any given observation time $t_0$
during the matter-dominated era, by rescaling down $\om_1 (t_0)$ to
$\om_0 $ as [13-15] $\om_1 (t_0)/\om_0 = k_1/k_0 \simeq (H_1 a_1)/(H_0
a_0)$, where $H_1=H(t_1)$, $a_1= a(t_1)$ and so on. By recalling that
$a\sim H^{-1/2}$ in the radiation phase, we obtain
$$
{\om_1 \over \om_0}=({H_1 \over H_2})^{1/2}
_{rad} ({H_2 a_2 \over H_0 a_0})
_{mat} =
({H_1 \over M_P})^{1/2} ({M_P a_P \over H_0 a_0}) \eqno(23)
$$
where the transition curvature scale  $H_1$ has been conveniently
expressed in units of Planck mass $M_P$. By introducing the black-body
temperature $T_\ga$, which evolves adiabatically ($a(t)T_\ga(t)=const$)
from $T_\ga (t_0)$ to $T_\ga (t_P)= M_P$, we find from eq.(23)
$$
\om_1(t_0)= ({H_1\over M_P})^{1/2} T_\ga (t_0) . \eqno(24)
$$
$S_g$ can thus be rewritten
$$
S_g\simeq |\da| S_\ga ({H_1\over M_P})^{3/2} \eqno (25)
$$
where $S_\ga \simeq [a(t) \ell T_\ga (t)]^3= const$ is the usual
black-body
entropy of the CMB radiation (in terms of the to-day parameters, $(
a_0\ell T_{\ga 0})^3 \sim (T_{\ga 0}/H_0)^3 \sim (10^{29})^3 $).

Eq.(25), which represent the main result of this paper, holds in general
for particle production with a spectrum $\overline n(\om) \sim \om ^{-
|\da|}$, quite irrespective of the kind of fluctuations which are
amplified (even for photons, in a suitable non-conformally flat
background), and is only weakly dependent on the background kinematics
(through the parameter $|\da|$, which is typically of order unity
[13-15]). Particle production from the vacuum is thus a process able to
explain the observed cosmological level of entropy, {\it provided the
curvature scale at the inflation-radiation transition is of the order of
the Planck one}, $H_1 \simeq M_P$.

In the standard inflationary scenario, based on an accelerated phase of
de Sitter-like exponential expansion, there are well known
phenomenological bounds [16] which imply $H_1\lsim 10^{-4} M_P$, and
which rule out the mechanism discussed here as a possible explanation of
the observed entropy. However, as recently stressed in [13,17,18], such
constraints are evaded if the de Sitter and/or the radiation-dominated
phase are preceeded by a phase of accelerated evolution and growing
curvature, like in the recently proposed "pre-big-bang" models
[18,19] of a duality-symmetric
string cosmology. We thus conclude that the result obtained in this
paper provides an additional reason of interest for such models, where
the maximum curvature scale is only constrained by [13,17,18]
$H_1 \lsim 10^2 M_P$.
\vskip 1.5 cm
{\bf Acknowledgements}

One of us (M.Giovannini) wishes to thank L.P.Grishchuk for an interesting
discussion. M.Gasperini is grateful to N.Sanchez and G.Veneziano for
discussions and helpful remarks.
\vskip 1.5 cm
{\bf Note added}

After this paper was submitted we received a preprint by R.
Brandenberger, T.Prokopec and V.Mukhanov, {\it "The entropy of the
gravitational field"} (Brown-HET-849, August 1992), whose content
overlap to some extent with ours, and in which a result similar
to our equation (25) is obtained through a different procedure.

\vfill\eject
\centerline{\bf References}

\item{1.}E.W.Kolb and M.S.Turner, The early universe (Addison-Wesley,
Redwood City, CA 1990), Chapt.8;

K.A.Olive, Phys.Rep.190(1990)307

\item{2.}E.W.Kolb, in Proc. of the 79th Nobel Symposium, Physica Scripta

T36(1991)199;

M.S.Turner, in Proc of the First Erice School "D.Chalonge" on

astrofundamental physics (September 1991), ed. by N.Sanchez

(World
Scientific, Singapore)

\item{3.}B.L.Hu and D.Pavon, Phys.Lett.B180(1986)329

\item{4.}B.L.Hu and H.E.Kandrup, Phys.Rev.D35(1987)1776;

H.E.Kandrup, Phys.Lett.B185(1987)382;

H.E.Kandrup, Phys.Lett.B202(1988)207;

H.E.Kandrup, J.Math.Phys.28(1987)1398

\item{5.}H.E.Kandrup, Phys.Rev.D37(1988)3505

\item{6.}L.P.Grishchuk and Y.V.Sidorov, Phys.Rev.D42(1990)3413;

L.P.Grishchuk, "Quantum mechanics of the primordial cosmological

perturbations", in Proc of the 6th Marcel Grossmann Meeting

(Kyoto, June
1991);

L.P.Grishchuk, "Squeezed states in the theory of primordial
gravitational

waves", in Proc. of the Workshop on squeezed states and
uncertainty

relations (Maryland Univ.), ed. by D.Han, Y.S.Kim and
W.W.Zachary

(Nasa Conf. Pub. No.3135, 1992) p.329

\item{7.}L.P.Grishchuk and Y.V.Sidorov, Class. Quantum Grav.6(1991)
L161;

L.P.Grishchuk, in Proc of the VIth Brazilian School on cosmology and

gravitation (Rio de Janeiro, July 1989)

\item{8.}J.Grochmalicki and M.Lewenstein, Phys.Rep.208(1991)189

\item{9.}B.L.Schumaker, Phys.Rep.135(1986)317

\item{10.}R.Brandenbeger, V.Mukhanov and T.Prokopec, "Entropy of a
classical

stochastic field and cosmological perturbations",
BROWN-HET-859 (June 1992)

\item{11.}T.Prokopec, "Entropy of the squeezed vacuum",
BROWN-HET-861

(June 1992)

\item{12.}M.Gasperini, Prog.Theor.Phys.84(1990)899

\item{13.}M.Gasperini and M.Giovannini, Phys.Lett.B282(1992)36;

"Dilaton contributions to the cosmic gravitational wave background",

DFTT-58/92 (to appear in Phys.Rev.D)

\item{14.}B.Allen, Phys.Rev.D37(1988)2078

\item{15.}V.Sanhi, Phys.Rev.D42(1990)453

\item{16.}V.A.Rubakov, M.V.Sazhin and A.V.Veryanskin, Phys.Lett.B115
(1982)189;

R.Fabbri and M.D.Pollock, Phys.Lett.B125(1983)445;

L.F.Abbot and M.B.Wise, Nucl.Phys.B244(1984)541

\item{17.}M.Gasperini and M.Giovannini, Class. Quantum Grav.9(1992)L137

\item{18.}M.Gasperini and G.Veneziano, "Pre-big-bang in string
cosmology", CERN TH.6572/92 (July 1992)

\item{19.}M.Gasperini, N.Sanchez and G.Veneziano, Nucl.Phys.B364(1991)
365;

G.Veneziano, Phys.Lett.B265(1991)287;

M.Gasperini, J.Maharana and G.Veneziano, Phys.Lett.B272(1991)277;

M.Gasperini and G.Veneziano, Phys.Lett.B277(1992)256;

G.Veneziano, in Proc of the 6th Int. Workshop on Theor.Phys.

"String
quantum gravity" (Erice, June 1992), ed. by N.Sanchez

(World Scientific,
Singapore)

\bye

\end